# A Comparison of High-Level Design Tools for SoC-FPGA on Disparity Map Calculation Example

Shaodong Qin, Mladen Berekovic
Chair for Chip Design for Embedded Computing, Technische Universität Braunschweig
D-38106 Braunschweig, Germany
Email: {qin, berekovic}@c3e.cs.tu-bs.de

*Abstract*—Modern SoC-FPGA that consists of FPGA with embedded ARM cores is being popularized as an embedded vision system platform. However, the design approach of SoC-FPGA applications still follows traditional hardware-software separate workflow, which becomes the barrier of rapid product design and iteration on SoC-FPGA. High-Level Synthesis (HLS) and OpenCL-based system-level design approaches provide programmers the possibility to design SoC-FGPA at system-level with an unified development environment for both hardware and software. To evaluate the feasibility of high-level design approach especially for embedded vision applications, Vivado HLS and Altera SDK for OpenCL, representative and most popular commercial tools in market, are selected as evaluation design tools, disparity map calculation as targeting application. In this paper, hardware accelerators of disparity map calculation are designed with both tools and implemented on Zedboard and SoCKit development board, respectively. Comparisons between design tools are made in aspects of supporting directives, accelerator design process, and generated hardware performance. The results show that both tools can generate efficient hardware for disparity map calculation application with much less developing time. Moreover, we can also state that, more directives (e.g., interface type, array reshape, resource type specification) are supported, but more hardware knowledge is required, in Vivado HLS. In contrast, Altera SDK for OpenCL is relatively easier for software programmers who is new to hardware, but with the price of more resources usage on FPGA for similar hardware accelerator generation.

## I. INTRODUCTION

Nowadays, SoC-FPGA, which consists of embedded CPUs (e.g., ARM), programmable logics and a rich set of peripherals, is being popularised as an embedded system platform in robot vision field. Zedboard and SoCKit development board are two typical SoC-FPGA boards supported by Xilinx and Altera, respectively. However, for a typical application targeting for SoC-FPGA platform, the hardware part and software part of the application have to be designed under different development environment with different programming languages, which becomes the barrier of its popularization.

Therefore, high-level design tools, which can design the application at system-level with high-level programming languages, are promoted by FPGA vendors, such as Vivado High-Level Synthesis (HLS) [1] provided by Xilinx, Altera SDK for OpenCL [2], Stratus HLS [3] by Cadence, Synphony C Compiler [4] by Synopsys, etc. Besides, many open source high-level design tools are proposed by academics as well, e.g., LegUP [5], GAUT [6], ROCCC [7], etc. According to the difference of supported source file, these design tools can be categorised into two classes: HLS tools and OpenCL-based high-level design tools.

### A. High-Level Synthesis Tools

Most of high-level design tools, which generate RTL code from C-based source code (e.g., C, C++, SystemC), can be categorised into HLS tools, such as Stratus HLS, Vivado HLS, ROCCC, LegUP, etc. HLS has been studied for years in research area and only recently starts to be utilised for real projects. LegUp is one of the best open source high-level synthesis tool being developed by scholars. LegUp framework allows researchers to improve C to Verilog synthesis without building an infrastructure from scratch [8]. However, presently, only few Altera FPGA boards are supported by LegUP compiler and ARM-accelerator hybrid synthesis mode is under development. Vivado HLS, released by Xilinx, is the most popular commercial HLS tool in market. It gives user the possibility to speedup IP creation by enabling C, C++ and System C specifications to be directly targeted into Xilinx FPGAs without the need to manually create RTL [1].

### B. OpenCL-based High-Level Design Tools

OpenCL-based high-level design, which employs OpenCL [9] as the programming language, is a relatively new methodology for application design on FPGA platform. OpenCL is an open standard for parallel programming of heterogeneous systems which can consist of CPU, GPU, DSP, FPGA, etc. For a typical SoC-FPGA-based embedded system, the system is usually composed of ARM core, memory and accelerators that generated by programming logics. If we consider the ARM core being the host and accelerators being the devices, the SoC system can be deemed to a heterogeneous system. Therefore, utilising OpenCL as the high-level design language for SoC-FPGA is also promising.

Altera SDK for OpenCL is the only commercial tool that supports OpenCL-based high-level design currently. It provides user a much faster and higher level software development flow for Altera FPGAs by abstracting away the traditional hardware FPGA development workflow [2]. Meanwhile, research that intends to explore open source framework for OpenCL on FPGA is also undergoing. For instance, in [10], an simple OpenCL framework is proposed to evaluate the interconnect implementation on FPGAs.

Recently, research works that utilise or evaluate high-level design tools emerged as well. In [11], a tri-diagonal matrix algorithm is implemented with VHDL, Altera SDK for OpenCL, and Vivado HLS, respectively, and then the generated hardware performance is compared. The possibility of utilising HLS tools in computational finance field are explored in [12]. LegUp, Altera SDK for OpenCL, Bluespec SystemVerilog [13], and Chisel [14] design tools are evaluated for database application accelerations purpose in [15]. All papers conclude that high-level design tools can dramatically shorten the developing time and generate proper hardware. However, all these work are targeting for CPU + FPGA platform. For SoC-FPGA platform which contains ARM + FPGA, no publish research result is available yet.

In order to evaluate development experience of high-level design tools and performance of generated hardware architecture on SoC-FPGA, especially targeting for embedded vision applications, disparity map calculation is selected as the evaluation application; Vivado HLS and Altera SDK for OpenCL, which are the most mature and commercialised tools available, as representative design tools; and Zedboard and SoCKit development board as the corresponding implementation platform, respectively.

The rest of this paper is structured as follows: Section II mainly introduces disparity map calculation application. Thereafter, the design and implementation process of disparity map calculation with these two tools are explained and compared in Section III. Finally, conclusions are drawn in Section IV.

## II. Disparity Map Calculation

Disparity map calculation, which also known as depth estimation, is a fundamental but important algorithm widely used in stereo vision systems.

For a stereo vision system, assuming a 3D world point $S$, its projected points on stereo images being $S_l(x,y)$ and $S_r(x',y)$, as shown in Figure 1. According to the geometry of stereo vision system, the axes of projected points $S_l(x,y)$ and $S_r(x',y)$ are different. To measure this axes difference, disparity value $d$ is introduced and can be computed by

$$d = x - x'. \qquad (1)$$

Assuming $S_l(x,y)$ is known, then the process of calculating disparity $d$ is actually to locate the matching point $S_r(x',y)$. In other words, disparity value calculation for each point is a stereo matching process. And the process of computing disparity value for all points between a pair of stereo images is disparity map calculation algorithm.

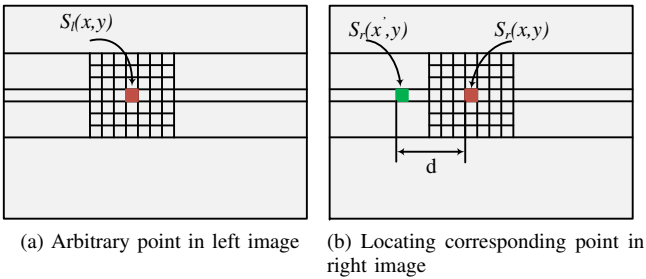

(a) Arbitrary point in left image  (b) Locating corresponding point in right image

Fig. 1: SAD-based stereo matching.

In order to accurately calculate disparity value for each pixel, various stereo matching algorithms, feature-based approaches (e.g., SIFT [16], SBM [11], etc.) and intensity-based approaches (e.g., SAD [17], SSD, etc.) are proposed. Here we don't want to explain in detail or compare different disparity map calculation algorithms since this is out of our topic. For the sake of simplicity, most typical Sum of Absolute Difference (SAD)-based stereo matching algorithm [17] is employed and implemented.

Referring to Figure 1, if we use a $N \times N$ floating window for each image with $S_l(x,y)$ and $S_r(x,y)$ being their respective centers, differences exist between two window blocks. However, if the center points being $S_l(x,y)$ and $S_r(x',y)$, respectively, little differences should exist. To measure these intensity differences, Sum of Absolute Difference (SAD) is introduced and calculated. If we denote $T_d(x,y)$ as the SAD of $S_l(x,y)$ and $S_r(x-d,y)$, $T_d(x,y)$ can be computed by

$$T_d(x,y) = \sum_{(i,j) \in N} |S_l(x+i,y+j) - S_r(x-d+i,y+j)|. \qquad (2)$$

Since the matching pixel in right image $S_r(x',y)$ should be on the left side of $S_l(x,y)$, pixel by pixel left shifting of the floating window is executed and the SAD is calculated for each shifting. Assuming the maximum disparity as $M$, after $M$ times shifting and computing, the minimum SAD $T(x,y)$ can be obtained by

$$T(x,y) = \min_{d \in M} T_d(x,y), \qquad (3)$$

and its corresponding disparity as the true disparity value at point $(x,y)$. Repeat this process for each pixel until disparity map of the whole image is obtained.

## III. Implementation on SoC-FPGA

After we are familiar with the SAD-based disparity map calculation algorithm, in this section, disparity map calculation accelerator is designed and implemented with both Vivado HLS and Altera SDK for OpenCL tools. The detailed implementation process is explained as follows.

### A. Accelerator Design with Vivado HLS

From the SAD-based disparity map calculation process description in Section II, we can notice that, for each pixel, $M$ times SAD operations should be executed and each SAD operation is composed of $N \times N$ pixels reading from each image, $N \times N$ times absolute difference calculation, and $N \times N - 1$ time absolute difference addition. It is obvious that too much memory accesses and computing logics are required for each pixel processing, which can be difficult for efficient hardware architecture generation on SoC. Therefore, some optimization techniques, especially for FPGA hardware implementation, are performed.

*1) Preserving Intermediate Results:* Originally, the absolute difference for each pixel-pair will be calculated $N \times N$ times. By preserving the absolute difference results, absolute difference operation for each pixel-pair only need to be executed once, therefore, much less computation resources are required. Meantime, simpler computation logic can simplify the pipeline architecture generated as well. The only price of

76

preserving intermediate results is that more local memories are required for intermediate data preservation. But comparing to the computation quantity that simplified, the cost is pretty small. In our application, column-SAD is the actual intermediate result being preserved, since it is difficult to explain column-SAD before describing the disparity map calculation process, it will be explained later in this section.

*2) Utilising Local Memory:* In disparity map calculation application, $N \times N$ pixels from each image should be read in in each pixel processing cycle, which is impossible due to the memory ports limitation. Utilising local memory, especially for image processing applications, which can store a few lines of image pixels, can dramatically improve the pixel accessing performance. The local memory size can be calculated by delicate hardware design. In our case, $N \times N$ floating window for each pixel pair is processed, if one pixel read in and one pixel processing is performed each cycle, $N$ lines pixels storage for each image are required. One pixel from each image is read in per cycle, when the fifth pixel of last row of local memory is read in, disparity map calculation process starts. If local memory is full-filled with pixels, new data will be stored from the first element of local memory again. If we denote the image width as $w$, the local memory size should be $2N \times w$. Besides, intermediate results can also be stored in local memory.

*3) Utilising Shifting Registers:* Utilising Shifting registers is an effective approach commonly used in FPGA design with Verilog/VHDL. Vivado HLS also provides the support for hardware generation shifting registers by simply adding directives. Shifting all registers left or right and then reading in new data guarantees that the latest data is always stored at the same location of register array. This can greatly simplify the indexing problem in local memory and is extremely useful for our col-SAD register table. Moreover, efficient/pipelined hardware architecture is easier to be generated with shifting registers utilisation.

*4) Other Techniques:* Some other techniques are deployed in our application as well, e.g., loop unrolling, pipeline declaration, resource core specification. Since these directives are simple and straightforward, the explanations will be omitted here.

With the techniques mentioned above, the disparity calculation processes as follows. Referring to Figure 2, $N \times w$ size local memory $LM$ and $RM$ are used as $N$-line-pixels buffers for left and right image, respectively. $LR$ is a $N$ size register array that used to store one column pixels which will be read from $LM$ local memory. $RR_i$ is an $M$ column registers table that each column can preserve one column pixels from $RM$ local memory, with $i = \{0, 1, \ldots, M-1\}$. col-SADs is an $M \times (N+1)$ size shifting registers table which is used to store column-SADs.

For each cycle, the right image buffer $RR_i$ and col-SADs perform one-pixel-left column-shifting. Thereafter, one pixel is read in from each image to local memory $LM$ and $RM$, respectively. And one column pixels are read in from local memory $LM$ and $RM$ to register array $LR$ and last column of register table $RR_i$, respectively. Column SADs are calculated between $LR$ and each column of $RR_i$. The results, according to different disparities, are stored into the last column of col-

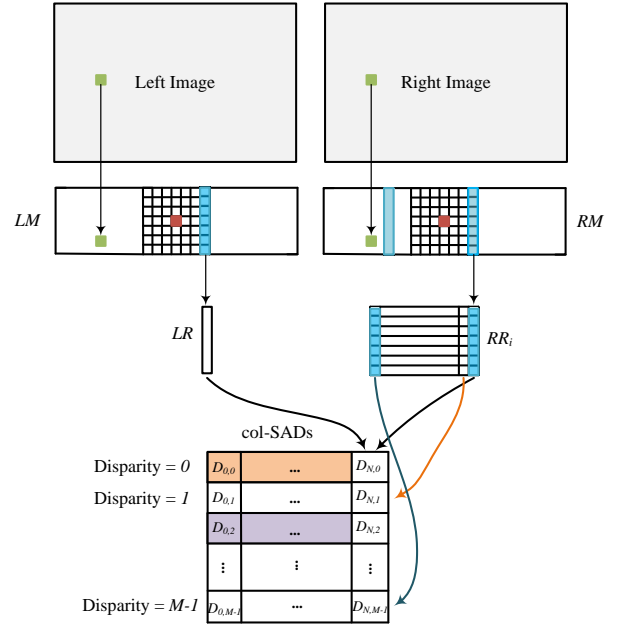

Fig. 2: Disparity map calculation process illustration.

SADs register table. For instance, the col-SAD of $LR$ and $RR_0$ is stored in the first element of last column of col-SADs $D_{N,0}$; $LR$ and $RR_{M-1}$ col-SAD result is stored in $D_{N,M-1}$; so on and so forth. Meanwhile, the first $N$ columns (0 to $N-1$) of col-SADs perform row sum, which is to get the SAD for each disparity. Finally the smallest sum corresponding disparity is selected as the real disparity at this point.

This approach reduces the memory access by using local memory and reduces computation by preserving col-SADs. Although, for every new line, all data in shifting registers need to be reloaded or recalculated. But since the computing process is pretty suitable for pipeline processing, efficient hardware architecture can be generated.

In this paper, 640×480 stereo images are used as input and Zedboard is deployed as the targeting SoC board. After we connect HLS generated accelerator with ARM core in Vivado, generate bitstream and on board testing, the results are finally obtained and listed in Table I. As Table I shows, different window size and max disparity settings are set up and the resources usages on Zedboard are listed as well. The reason we select these setups is in order to make the comparison with Altera SDK for OpenCL generated hardware. Because many shifting registers and computing resources are utilised in the algorithm, therefore, a lot of logics are used on Zedboard. As for the memory usage, only two $N$-line-pixels are stored in local memory, which is relatively little portion on Zedboard. The last column of Table I lists the running FPS of disparity map calculation accelerator on Zedboard. It is obvious that FPS 228 doesn't really change with different setups. The reasonable explanation for that is deep pipelined hardware is generated for each setup, which means one pixel processing each clock cycle is truly executed. Therefore, for the same resolution rate stereo images, similar running FPS should be obtained on board.

| Setup | | Resource Usage | | | FPS |
|---|---|---|---|---|---|
| Window Size | Max Disparity | Logic | Mem | DSP | |
| 7 × 7 | 80 | 60.8% | 5% | 0% | 228 |
| 9 × 9 | 64 | 59.8% | 6% | 0% | 230 |
| 9 × 9 | 78 | 83.3% | 6% | 0% | 228 |
| 9 × 9 | 80 | 84.6% | 6% | 0% | 228 |

TABLE I: Results of disparity map calculation with Vivado HLS generated accelerator.

| Setup | | Resource Usage | | | FPS |
|---|---|---|---|---|---|
| Window Size | Max Disparity | Logic | Mem | DSP | |
| 7 × 7 | 80 | 75% | 23% | 0% | 242 |
| 9 × 9 | 64 | 82% | 29% | 0% | 208 |
| 9 × 9 | 70 | 89% | 29% | 0% | 198 |
| 9 × 9 | 78 | 98% | 29% | 0% | 193 |

TABLE II: Results of disparity map calculation with Altera SDK for OpenCL generated accelerator.

*B. Accelerator Design with Altera SDK for OpenCL*

Unlike C++ as our application source file type in Vivado HLS, Altera SDK for OpenCL utilises OpenCL as its unified programming language for both hardware and software, therefore, some OpenCL framework setup operations, such as platform setup, memory setup, etc., should be added into the original source file first. However, before touching the kernel code with OpenCL, one thing we should determine is which kind of kernel structure shall be deployed: single work-item kernel or NDRange kernel.

Altera SDK for OpenCL supports two different types of kernel, NDRange kernel and single work-item kernel. Single work-item kernel can also be seen as a $(1, 1, 1)$ NDRange kernel. But for Altera SDK for OpenCL, different directives and compiler behaviours are supported for different type of kernel. Therefore, proper kernel type should be deployed according to application characteristics. For the application with lots of data dependency or many loops, single work-item kernel can generate efficient pipelined hardware architecture and boost up the performance. On the contrary, if little data dependency or loops exist in the application, NDRange kernel can execute multiple Processing Elements (PEs) in parallel and improve the processing efficiency [18]. In our case, according to the calculation process in Section II, nested loops and data dependency exist. Therefore, single work-item kernel should be a better choice for our application. As a matter of fact, for single work-item kernel, the kernel code is similar to the original C code, therefore, only some specifications/directives are required.

Since the application is already well optimised with some techniques and directives in Section III-A, similar techniques are used in Altera design tool version as well. However, unlike Vivado HLS, Altera SDK for OpenCL doesn't provide any assistant window to help with the directive addition, therefore, all the directives must be added manually according to the user guide document [18] [19]. Thereafter, setting up same parameters (window size and max disparity) as in Vivado HLS, the whole system (hardware and software) is generated for SoCKit Development board. Unlike Vivado HLS, no further design or processing is required for embedded system generation. After on board testing with generated hardware architecture, the resource usage and performance results are obtained and listed in Table II.

From Table II, we can state that, efficient hardware accelerator can be generated with Altera SDK for OpenCL as well. But due to the resource limitation on SoCKit, which is actually similar to Zedboard, with $9 \times 9$ window size setting-up, only 78 maximum disparity can be achieved. Moreover, Table II also shows that, comparing to Vivado HLS generated hardware, around 10% more programming logics are use. As for memory usage, in OpenCL programming model, the whole stereo images are stored in the allocated global memory, therefore, much more memory is and should be utilised.

*C. Comparison*

After we design and implement disparity map calculation application with Vivado HLS and Altera SDK for OpenCL, respectively, we can conclude that, both tools can generate efficient hardware architecture with much shorter developing time. But the tools themselves and development experiences with them are actually quite different. Vivado HLS is a tool that used to speedup hardware accelerator (IP) design and creation on FPGA. The generated file is IP core, which can be connected to other IP or ARM in Vivado system design. Therefore, some work should be done in Vivado as well and hardware knowledge is necessary for the user. Besides, more directives, such as memory type, interface, loop control, etc., are supported in Vivado HLS. By contrast, Altera SDK for OpenCL is a tool provided for software programmers, which means little hardware knowledge is required for user and much easier workflow for the whole process. However, comparing to Vivado HLS, less directives are supported and more difficult to debug applications. Of course, OpenCL knowledge is required for the programmer.

IV. CONCLUSION

This paper explores high-level design approaches on SoC-FPGA platform with disparity map calculation as the targeting application. Vivado High-Level Synthesis (HLS) and Altera SDK for OpenCL, two representative and most mature commercial tools, are selected as the design tools. Zedboard and SoCKit development board are the corresponding implementation SoC-FPGA, respectively. Hardware accelerators of disparity map calculation are designed with Vivado HLS and Altera SDK for OpenCL, respectively, and comparisons are made for the implementation process and generated hardware performance. From the comparison, we conclude that, both design tools can generate efficient hardware for disparity map calculation application with much less developing time. For $640 \times 480$ stereo images, window size being $9 \times 9$, and max disparity being 80 (78 in Altera SDK for OpenCL), 228 and 193 fps disparity map calculation can be achieved, respectively. However, for same algorithm and setups, more resources on FPGA are used for the accelerator implemented by Altera SDK for OpenCL. Besides, more hardware knowledge is required for Vivado HLS user and Altera SDK for OpenCL is more suitable for software programmers.